\documentclass[aps,twocolumn,nofootinbib,superscriptaddress]{revtex4-1}
\usepackage{amsfonts,ulem,bbold}
\usepackage{amssymb,amsmath}%,graphicx,wrapfig,epsfig}
\usepackage{lipsum}
\usepackage{color}
\newcommand{\nn}{\nonumber}

\newcommand{\be}{\begin{equation}}  
\newcommand{\ee}{\end{equation}}  
\newcommand{\bea}{\begin{eqnarray}}  
\newcommand{\eea}{\end{eqnarray}}

\newcommand{\Comment}[1]{{}}
\definecolor{MyDarkBlue}{rgb}{0.15,0.15,0.45}
\usepackage[linktocpage=true]{hyperref}
\hypersetup{
colorlinks=true,
citecolor=MyDarkBlue,
linkcolor=MyDarkBlue,
urlcolor=MyDarkBlue,
}

\begin{document}

\title{Positivity Constraints for Pseudo-linear Massive Spin-2 and Vector Galileons} 

\author{James Bonifacio}
\affiliation{Theoretical Physics, University of Oxford, DWB,\\
Keble Road, Oxford, OX1 3NP, UK}  
\author{Kurt Hinterbichler}
\affiliation{Perimeter Institute for Theoretical Physics,\\
31 Caroline St. N, Waterloo, Ontario, Canada, N2L 2Y5}  
\author{Rachel A. Rosen}
\affiliation{Physics Department and Institute for Strings, Cosmology, and Astroparticle Physics,\\
Columbia University, New York, NY 10027, USA}

\begin{abstract}
We derive analyticity constraints on a nonlinear ghost-free effective theory of a massive spin-2 particle known as pseudo-linear massive gravity, and on a generalized theory of a massive \mbox{spin-1} particle, both of which provide simple IR completions of Galileon theories.  For pseudo-linear massive gravity we find that, unlike dRGT massive gravity, there is no window of parameter space which satisfies the analyticity constraints.  
For massive vectors which reduce to Galileons in the decoupling limit, we find that no two-derivative actions are compatible with positivity, but that higher derivative actions can be made compatible.

\end{abstract}  
  
\maketitle

\noindent
{\section{Introduction and Summary}}
\vspace{-.2cm}
In recent years great progress has been made in constructing and understanding effective field theories of massive spin-2 particles, i.e. theories of massive gravity.  Most notably, the construction of de Rham, Gabadadze and Tolley (dRGT) \cite{deRham:2010kj} has been shown to be free of the extra degree of freedom known as the Boulware-Deser ghost \cite{Boulware:1973my,Hassan:2011hr} (see \cite{Hinterbichler:2011tt,deRham:2014zqa} for reviews). The dRGT theory consists of the Einstein-Hilbert term plus a specific two-parameter potential.  It is a low energy effective field theory with a strong coupling scale $\Lambda_3 \equiv (m^2M_p)^{1/3}$, where $m$ is the graviton mass, and $M_p$ is the Planck mass appearing in front of the Einstein-Hilbert term. 

An outstanding question is whether or not dRGT massive gravity can be UV completed.  Completion could require the addition of new heavy degrees of freedom, or the theory may already be complete but require an understanding of strongly coupled quantum effects at high energies (i.e., asymptotic safety). 

However, there are results in the literature that would seem to suggest that a local Lorentz invariant UV completion of massive gravity is not possible.  In general, the requirements of S-matrix analyticity and unitarity in a local and Lorentz invariant UV theory puts constraints on the values of certain couplings in the low energy theory.  The simplest such constraints come from analyticity requirements on the four-particle amplitude in the forward limit (see, e.g., \cite{Adams:2006sv}).  When these constraints are applied to a specific scalar theory with an enhanced shift symmetry known as the Galileon \cite{Nicolis:2008in}, the theory is found to lie right on the border of the region inside of which these constraints are satisfied, and thus marginally fails the constraints \cite{Adams:2006sv}. (However, there are subtleties with the assumptions involved when these constraints are applied to massless particles such as the Galileon.)  In a high energy limit where the strong coupling scale is held fixed, known as the decoupling limit, the helicity-0 mode of the massive graviton behaves like the Galileon\footnote{More precisely, there is a one parameter family of interactions for which the decoupling limit is a Galileon, in the other cases it is a scalar-tensor generalization of the Galileon \cite{deRham:2010ik}.}.   Thus it might be assumed that dRGT massive gravity inherits the same issues with analyticity constraints.

Recently, in a very interesting paper \cite{Cheung:2016yqr}, it was shown that in the full dRGT massive gravity, interactions present away from the decoupling limit can push the theory away from this border and off to either side, depending on the choice of parameters.  A window of parameter space opens up in which the analyticity constraints are satisfied.  Within this window dRGT massive gravity can be thought of as an ``IR completion" of the Galileon which may allow for a standard Lorentz invariant UV completion.  

In this paper we explore other IR completions of the Galileon and their analyticity constraints.  In fact, dRGT massive gravity is not the unique ghost-free interacting massive spin-2 with a $\Lambda_3$ strong coupling scale.  There is also the possibility of ``pseudo-linear" massive gravity \cite{Hinterbichler:2013eza}, which does not have Einstein-Hilbert interactions.  Pseudo-linear massive gravity has three interactions terms.  Two of these are potentials, and have a decoupling limit identical to that of dRGT, so this theory provides an alternative IR completion of the Galileon.  The third interaction term is a non-Einstein two derivative cubic interaction \cite{Folkerts:2011ev}, which has no known dRGT counterpart \cite{deRham:2013tfa,Kimura:2013ika,deRham:2015rxa}.

The pseudo-linear theory has a much simpler structure than the dRGT theory.  In particular, it does not have the complicated non-linearities of the Einstein-Hilbert term.  Because of this relative simplicity, it might be easier to find a UV completion for the pseudo-linear theory, should it satisfy the analyticity constraints.

Here, we compute four-particle forward limit analyticity constraints for the three parameter pseudo-linear massive gravity.  We find that, unlike the case of dRGT massive gravity, there is no window of parameter space for which the positivity constraints are obeyed.

As another example of an IR completion of the Galileon, we also consider a self-interacting massive vector boson.  In particular, we focus on a massive vector whose helicity-0 mode gives a Galileon theory in the decoupling limit.  In this case, we find that no two-derivative Lagrangian satisfies positivity, but with the addition of higher derivative terms, positivity can be satisfied.

It is important to keep in mind what the analyticity constraints tell us.  The analyticity constraint we are discussing is a constraint on the four-particle S-matrix in the forward limit.  Satisfying this constraint is a necessary but not sufficient condition for UV completion by a local Lorentz invariant quantum field theory.  It is not sufficient because even if this forward limit four-particle constraint is satisfied, there could always be further constraints on higher point amplitudes, at different kinematics, or other considerations entirely, that are not satisfied.   On the other hand, if these constraints are violated, it does not mean that the effective field theory is inconsistent.  It simply means that the UV completion, if it exists, is not a local Lorentz invariant field theory of the usual type.   It could be something more unusual; perhaps not strictly local (for example the proposal of \cite{Keltner:2015xda} for the Galileon, or \cite{Cooper:2013ffa} for wrong sign DBI in two dimensions), not Lorentz invariant, or not a field theory at all.  

\vskip.4cm

\noindent
{\section{Pseudo-linear Massive Gravity}}

\vspace{-.2cm}

In this section we review the pseudo-linear theory of massive gravity.  For further details, see \cite{Hinterbichler:2013eza}.  The quadratic part of the Lagrangian is the Fierz-Pauli Lagrangian \cite{Fierz:1939ix} describing a pure spin-2 particle of mass $m$ using a symmetric tensor field $h_{\mu\nu}$,
\bea
\label{FP}
\begin{array}{lcl}
{\cal L}_{FP}&=& -\frac{1}{2} \partial_\lambda h_{\mu\nu}\partial^{\lambda}h^{\mu\nu}+\partial_\mu h_{\nu \lambda}\partial^\nu h^{\mu\lambda}-\partial_\mu h^{\mu\nu}\partial_\nu h \vspace{0.2cm} \\
&& + \frac{1}{2}\partial_\lambda h \partial^\lambda h -\frac{1}{2} m^2(h_{\mu\nu}h^{\mu\nu}-h^2) \, .
\end{array}
\eea
In four dimensions there are three possible pseudo-linear terms we can add to \eqref{FP} such that the total number of degrees of freedom are unchanged,
\be
\label{pseudolin}
{\cal L} = {\cal L}_{FP} +\frac{1}{M_p}\lambda_1 {\cal L}_{2,3} + \frac{m^2}{M_p}\lambda_3 {\cal L}_{0,3} + \frac{m^2}{M_p^2} \lambda_4{\cal L}_{0,4} \, .
\ee
Here $\lambda_1,\lambda_3,\lambda_4$ are dimensionless couplings and $M_p$ is a mass scale suppressing powers of the field.  (One of the $\lambda$'s is redundant and can be absorbed into $M_p$, but we will keep all three explicit so as to keep track of signs and when coefficients vanish.)
The first of these terms ${\cal L}_{2,3}$ is a two-derivative cubic interaction that is {\it not} the cubic truncation of Einstein-Hilbert,
\be
\label{L23}
{\cal L}_{2,3} = 12\, \delta^{[\mu_1}_{\nu_1} \delta^{\mu_2}_{\nu_2} \delta^{\mu_3}_{\nu_3} \delta^{\mu_4]}_{\nu_4}\, (\partial_{\mu_1}\partial^{\nu_1}  h_{\mu_2}^{~\nu_2})\,   h_{\mu_3}^{~\nu_3}\, h_{\mu_4}^{~\nu_4} \, ,
\ee
where we anti-symmetrize with weight one.
The latter two terms ${\cal L}_{0,3}$ and ${\cal L}_{0,4}$ are symmetric polynomials in $h_{\mu\nu}$,
\be
\label{L03}
{\cal L}_{0,3} = \frac{1}{6} \left([h]^3-3[h][h^2]+2[h^3]  \right) \, ,
\ee
\be
\label{L04}
 {\cal L}_{0,4} = \frac{1}{24} \left([h]^4-6[h]^2[h^2]+3[h^2]^2+8[h][h^3]-6[h^4]  \right) \, .
\ee
We have used notation $h$ for the matrix $h^\mu_{\ \nu}$ where the upper index is raised with the Minkowski metric.  The brackets indicate traces of the enclosed matrix products. 
 
The overall scaling on $m$ and $M_p$ is chosen so that one indeed recovers a Galileon theory for the helicity-0 mode of the massive graviton in the limit  $m \rightarrow 0$ and $M_p \rightarrow \infty$ with $\Lambda_3 \equiv (m^2M_p)^{1/3}$ fixed.  Specifically, just as in dRGT massive gravity, the zero-derivative terms generate the Galileon terms for the helicity-0 mode in this ``decoupling" limit, as well as a mixing term between the helicity-0 and helicity-2 modes.  Differing from dRGT, the derivative interaction ${\cal L}_{2,3}$ gives a new contribution to the scalar-tensor sector in the decoupling limit.

\vskip.4cm

\noindent
{\section{Constraints}}

\vspace{-.2cm}

To derive constraints on the parameters  $\lambda_1$, $\lambda_3$, and $\lambda_4$, we follow the procedure of \cite{Cheung:2016yqr} and refer readers there for further details (see also \cite{Bellazzini:2016xrt}).  We consider the four-particle amplitude ${\cal A}(s,t)$ for the scattering of some crossing-symmetric choice of polarizations of our spin-2 particle.  This amplitude is that of an assumed UV complete theory whose low energy limit is our pure spin-2 effective field theory.   We take the forward limit $t\rightarrow 0$ and consider the quantity $f$ defined by the contour integral
\be
\label{f}
f=\frac{1}{2\pi i} \oint_\Gamma ds \frac{{\cal A}(s,0)}{(s-\mu^2)^3} \, .
\ee
Here $\mu$ is an arbitrary mass scale taken to be in the interval $0<\mu^2<4m^2$.  The contour $\Gamma$ is chosen such that it encircles the single particle poles $s=m^2$, $s=3m^2$ and $s=\mu^2$.  The single particle poles come from tree level exchange in the effective theory, and so their residues and hence $f$ can be computed at tree level solely within the effective theory.   Given unitarity and locality and the fact that our theory has a mass gap, the Froissart bound applies \cite{Froissart:1961ux,Martin:1962rt} and one can deform the contour to encircle the multi-particle branch cuts starting at $s=4m^2$ and $s=0$, dropping the vanishing boundary contour at infinity.  By the optical theorem and crossing symmetry, the value obtained from encircling the branch cuts is related to the total cross section which must be positive, so one finds that \eqref{f} must be strictly positive for a theory with any non-trivial interaction: 
\be f>0\, .\ee

We proceed to calculate $f$ for the pseudolinear theory \eqref{pseudolin}.   The amplitude goes as ${\cal A}(s,0) \sim s^2$ in the forward limit, as it does in dRGT, and it is these $s^2$ parts of the amplitude which are constrained by positivity.  We will only need to consider the scattering of external polarizations that are purely tensor (T), vector (V) or scalar (S).  As for dRGT massive gravity, the result is independent of the arbitrary mass scale $\mu$.  We find:
\bea
\begin{array}{rcl}
f(TTTT)_+ &=& \dfrac{9\lambda_1^2+4 \lambda_1\lambda_3}{3\, m^2 M_p^2}\vspace{0.2cm} \, ,\\
f(TTTT)_- &=& \dfrac{\lambda_1^2}{m^2 M_p^2} \vspace{0.2cm}\, ,\\
f(TVTV) &=& -\dfrac{3\lambda_1^2+4\lambda_1\lambda_3}{16\, m^2 M_p^2} \vspace{0.2cm}\, ,\\
f(TSTS) &=& - \dfrac{4\lambda_1^2+2\lambda_1 \lambda_3}{3\, m^2 M_p^2} \vspace{0.2cm}\, ,\\
f(VVVV)_+ &=& - \dfrac{15\lambda_1^2 +13\lambda_1\lambda_3+ 5 \lambda_3^2 }{12\, m^2 M_p^2} \vspace{0.2cm}\, ,\\
f(VVVV)_- &=& - \dfrac{15\lambda_1^2+4\lambda_1\lambda_3 -4 \lambda_3^2 + 4 \lambda_4}{16\, m^2 M_p^2} \vspace{0.2cm}\, ,\\
f(VSVS) &=&  -\dfrac{3\lambda_1^2 -8 \lambda_1\lambda_3 -12 \lambda_3^2 +8 \lambda_4}{48\, m^2 M_p^2} \vspace{0.2cm}\, ,\\
f(SSSS) &=& - \dfrac{5\lambda_1^2 + 6 \lambda_1\lambda_3+ \lambda_3^2 + 2 \lambda_4}{9\, m^2 M_p^2} \vspace{0.2cm}\, .\\
\end{array}
\eea
Here we refer to \cite{Cheung:2016yqr} for notation and conventions of the polarizations.  In particular, the ``$+$" and ``$-$" subscripts correspond to choosing polarizations that are parallel and orthogonal respectively.

We can see from these results that there is no choice of parameters that satisfies positivity bounds for all helicities.  From $f(TTTT)_-$ we see that we must have $\lambda_1\not=0$.  We may then scale the Planck mass, and flip the sign of $h_{\mu\nu}$ if necessary, to set $\lambda_1=1$.  Putting this into $f(VVVV)_+$, the resulting polynomial in $\lambda_3$ never achieves a positive value.  Thus, unlike dRGT massive gravity, there is no window of parameter space consistent with positivity constraints.

\vskip.4cm

\noindent
{\section{Massive Vector}}
\vspace{-.2cm}
It is natural to ask whether other theories, perhaps even simpler ones, can provide an IR completion of the Galileon consistent with positivity.  For instance, one might wonder if the tensor modes of the massive graviton are needed or if a window of positivity can be found for a theory that contains only massive vector modes.  Theories of ghost-free massive vectors have attracted interest recently \cite{Heisenberg:2014rta,Hull:2014bga,Tasinato:2014mia,Allys:2015sht,Hull:2015uwa,Charmchi:2015ggf,Jimenez:2016isa,Allys:2016jaq}, and within this class there are examples where the longitudinal mode behaves like a Galileon in the massless limit.

Here, we consider the most general two-derivative theory of a massive vector whose helicity-0 mode has Galileon interactions in the decoupling limit,
\bea
\label{vec}
\begin{array}{rcl}
{\cal L} &=& -\frac{1}{4} F_{\mu\nu}^2+m^2 A^2 +\lambda_3 \frac{m}{M_p}A^\mu A^\nu \partial_\mu A_\nu \\
&&+\lambda_4 \frac{1}{M_p^2} A^2(\partial_\mu A^\mu \partial_\nu A^\nu-\partial_\mu A^\nu\partial_\nu A^\mu) \\
&&+\lambda_5 \frac{1}{M_p^2} A_\mu A_\nu \partial_\lambda A^\mu (\partial^\lambda A^\nu - \partial^\nu A^\lambda) \\
&&+\lambda_6 \frac{1}{M_p^2} A^2 \partial_\mu A_\nu (\partial^\mu A^\nu - \partial^\nu A^\mu)\\
&&+\lambda_7 \frac{1}{M_p^2} A_\mu A_\nu( \partial^\mu A_\lambda \partial^\nu A^\lambda - \partial_\lambda A^\mu\partial^\lambda A^\nu)\\
&&+{\cal O}\left(A^5\right) \, , \label{vecgenactn}
\end{array}
\eea
with $F_{\mu\nu}\equiv \partial_\mu A_\nu-\partial_\nu A_\mu$.  We only need terms up to fourth order in the fields because we are only considering constraints from tree-level four-particle scattering.

To see the Galileon limit, we extract the behavior of the helicity-0 mode of the massive vector by making the replacement,
\be
A_\mu \rightarrow A_\mu +\frac{1}{m}\partial_\mu \phi \,,
\ee
and then taking the limit  $m \rightarrow 0$ and $M_p \rightarrow \infty$ with $\Lambda_3 \equiv (m^2M_P)^{1/3}$ fixed,
\bea
{\cal L}_{\rm DL} = -\frac{1}{4} F_{\mu\nu}^2+(\partial \phi)^2 +\lambda_3 \frac{1}{m^2 M_p}\partial^\mu \phi \partial^\nu \phi \partial_\mu \partial_\nu \phi \nonumber\\
+\lambda_4 \frac{1}{m^4 M_p^2} (\partial \phi)^2\left((\square\phi )^2-\partial_\mu \partial_\nu\phi\partial^\nu \partial^\mu\phi\right) \, .  \nn\\ \label{vecdecouplinglimlagn}
\eea
This is the action for a cubic and quartic Galileon \cite{Nicolis:2008in}.  Thus the scattering amplitude for the longitudinal mode goes like $\sim {s^3\over m^4 M_P^2}$ at high energy, just as in dRGT massive gravity, but will acquire an ${\cal O}(s^2)$ part beyond the decoupling limit which will be constrained by the positivity analysis.

We perform the same analysis as above for the massive vector Lagrangian \eqref{vec} and consider scattering  for pure vector and scalar states.  We find
\bea
\begin{array}{rcl}
f(VVVV)_\pm &=& 0 \vspace{0.2cm}\, ,\\
f(VSVS) &=& \, \dfrac{\lambda_7}{m^2 M_p^2} \vspace{0.2cm}\, ,\\
f(SSSS) &=& -\dfrac{\lambda_3^2+2\lambda_5-4 \lambda_7}{m^2 M_p^2} \vspace{0.2cm}\, .
\end{array} \label{fvaluesvecn}
\eea
From this we see that, while the components involving the scalar polarization can be made positive, no choice of coefficients is consistent with $f(VVVV)_\pm>0$.

The values of $f$, however, are sensitive to higher derivative operators which we have neglected by restricting to a  two-derivative action in \eqref{vecgenactn}.  And, unlike the case of pseudo-linear massive gravity, such terms can be included without introducing extra degrees of freedom into the theory.  For example, the following gauge invariant 4-derivative operators familiar from the Euler-Heisenberg Lagrangian,
\be {1\over m^2 M_p^2}\left[c_1 F^\mu_{\ \nu}F^\nu_{\ \rho}F^\rho_{\ \sigma}F^\sigma_{\ \mu}+c_2(F_{\mu\nu}F^{\mu\nu})^2\right]\, ,\ee
do not affect the decoupling limit \eqref{vecdecouplinglimlagn} nor do they introduce additional degrees of freedom.  But they do add the contributions 
\bea
\begin{array}{rcl}
 f(VVVV)_+ &=& {8\over m^2 M_p^2 }(c_1+2c_2)\, , \\
 f(VVVV)_- &=& {4\over m^2 M_p^2 }c_1\, , 
 \end{array}
 \eea
to \eqref{fvaluesvecn}, which now allows for positivity for the definite helicity scattering.  

In fact, one can show that certain choices of parameters can give $f > 0$ for any linear combination of polarizations and not just definite helicity scattering. For instance, one can make the following parameter choice:
\bea
\begin{array}{l}
c_1={1\over 4} ,  \,c_2=-{1\over 16} \, ,   \,\lambda_3=1 \, \vspace{0.2cm} \\
\lambda_4={1\over 2}, \, \lambda_5=1,\, \lambda_6=0, \, \lambda_7=1\, ,
\end{array}
 \eea
which gives 
\bea
\label{fmassvec}
\begin{array}{rcl}
f &=&\,  \frac{(vV_1^2 + vV_2^2+vS^2 )(wV_1^2 + wV_2^2+wS^2 )}{m^2 M_p^2}\, ,
\end{array}
\eea
where $vV_1$, $vV_2$ and $vS$ are respectively the coefficients of the two vector and one scalar polarizations of the first incoming particle and $wV_1$, $wV_2$ and $wS$ are respectively the coefficients of the two vector and one scalar polarizations of the second incoming particle.  As can be seen from \eqref{fmassvec}, $f$ is a sum of squares for this choice of parameters and is thus always positive regardless of the choice of polarizations.

\vskip.4cm

\noindent
{\section{Conclusions} }  
\vspace{-.2cm}

We have derived positivity constraints on the four-particle amplitude of the pseudo-linear massive spin-2 theory and two-derivative generalized Proca theories with Galileon decoupling limits.  We have shown that these simple ways of embedding the Galileon into an IR theory do not have the same success as dRGT massive gravity when it comes to obeying positivity constraints.  However, in the case of the generalized Proca theory, these constraints can be satisfied with the inclusion of higher derivative terms like those of the Euler-Heisenberg Lagrangian.

The failure of certain theories to obey positivity constraints does not mean they are necessarily inconsistent: it means that a UV completion, if it exists, must violate either strict locality, Lorentz invariance, unitarity, or some other assumption usually made in quantum field theory, and so must be something more exotic.  In contrast, it is possible that dRGT massive gravity may possess an ordinary, local and Lorentz invariant UV completion, at least within the window of parameter space identified in \cite{Cheung:2016yqr}.  Within the class of known ghost-free non-linear massive spin-2 theories, our results suggests that the presence of the Einstein-Hilbert interactions are necessary for such a completion. \\

\vspace{.5cm}

\noindent
{\bf Acknowledgements:} We are grateful to Cliff Cheung and Grant Remmen for helpful comments on the draft.  Research at Perimeter Institute is supported by the Government of Canada through Industry Canada and by the Province of Ontario through the Ministry of Economic Development and Innovation. This work was made possible in part through the support of a grant from the John Templeton Foundation. The opinions expressed in this publication are those of the author and do not necessarily reflect the views of the John Templeton Foundation (KH).  JB is supported by the Rhodes Trust.  RAR is supported by DOE grant DE-SC0011941.

\bibliographystyle{apsrev4-1}
\bibliography{pseudo5}
\end{document}